\documentclass[traditabstract]{aa}
\usepackage{graphicx}
\usepackage{natbib}
\usepackage{txfonts}
\bibpunct{(}{)}{;}{a}{}{,}
\begin{document}

  \title{The Horizontal Branch of the Sculptor Dwarf galaxy}
  \author{Maurizio Salaris\inst{1},  
          Thomas de Boer\inst{2}, Eline Tolstoy\inst{3},  
          Giuliana Fiorentino\inst{4} \and 
          Santi Cassisi \inst{5}}

\institute{Astrophysics Research Institute, 
           Liverpool John Moores University
           IC2, Liverpool Science Park
           146 Brownlow Hill
           Liverpool L3 5RF, UK\\
           \email{M.Salaris@aljmu.ac.uk}         
           \and
           Institute of Astronomy, University of Cambridge, Madingley Road, Cambridge, UK, CB3 0HA
           \and
           Kapteyn Astronomical Institute,
           University of Groningen, 
           Postbus 800, 9700 AV Groningen, 
           The Netherlands
           \and
           Dipartimento di Fisica e Astronomia, Università degli Studi di Bologna, 
           Viale Berti Pichat 6/2, I$-$40127 Bologna, Italy          
           \and
           INAF~$-$~Osservatorio Astronomico di Collurania, Via M. Maggini, I$-$64100 , Teramo, Italy
           }
  \abstract
  {We have performed the first detailed simulation of the horizontal branch of the Sculptor dwarf spheroidal galaxy 
by means of synthetic modelling techniques, 
taking consistently into account the star formation history and metallicity evolution as determined from the main sequence 
and red giant branch spectroscopic observations. The only free parameter in the whole analysis is the 
integrated mass loss of red giant branch stars.
This is the first time that synthetic horizontal branch models,  
consistent with the complex star formation history of a galaxy, are calculated and matched to the observations.
We find that the metallicity range covered by the star formation history, 
as constrained by observations, 
plus a simple mass loss law, enable us to cover both the full magnitude and colour range of HB stars.  
In addition the number count distribution along the observed horizontal branch, can be 
also reproduced, provided that the red giant branch mass loss 
is mildly metallicity dependent, with a very small dispersion at fixed metallicity. 
The magnitude, metallicity and period distribution of the RR Lyrae stars are also well reproduced.
There is no excess of bright objects that require enhanced-He models. The lack of 
signatures of enhanced-He stars along the horizontal branch is consistent with the lack of the O-Na anticorrelation 
observed in Sculptor and other dwarf galaxies, and confirms the intrinsic difference between Local Group dwarf galaxies and globular 
cluster populations. We also compare the brightness of the observed red giant branch bump with the synthetic counterpart, and find 
a discrepancy -- the theoretical bump being brighter -- similar to what is observed in Galactic globular clusters.}
\authorrunning{M. Salaris et al.}
  \maketitle

%_____________________________________________________________________

\section{Introduction}

The Horizontal Branch (HB) is a luminous feature seen in the Colour-Magnitude Diagrams (CMDs) of all galaxies and 
globular clusters, corresponding to the central He-burning phase of low mass stars. 
Its shape -- in a given bandpass combination -- 
varies significantly, with an obvious dependence upon the age and metallicity range of the stellar populations 
of the system. However, several detailed investigations of the HB morphology in 
globular clusters have since long demonstrated that age and metallicity alone do not allow a unique interpretation of 
the HB \citep[see, e.g.,][for a recent study, and references therein]{dotterHB}.

The most critical problem is that it is not possible to predict from first principles the mass loss during the 
Red Giant Branch (RGB) phase, which determines the location along the HB for a star with a fixed initial mass and 
chemical composition. Other factors, such as the initial helium abundance, also appear to play a crucial role in determining 
the form of the observed HB in Galactic globulars 
\citep[see, e.g.,][and references therein]{HBsp, danto, ema}. The role played by He appears to be related 
to the recently emerged new picture of the globular cluster origin and evolution, whereby each cluster hosts 
first generation (FG) stars with {\sl normal} He 
(mass fraction roughly equal to the cosmological He abundance) and $\alpha$-enhanced metal mixtures, 
and second generation objects with C-N, O-Na (and sometimes Mg-Al) anticorrelations and 
enhanced He \citep[see, e.g.][and references therein for a review]{gratton}. 

So far both these critical parameters -- RGB mass loss and initial He of individual stars  -- have proven hard to measure 
\citep[see, i.e.,][for some determinations on a small 
sample of globular clusters]{origlia, villanova, villanova_b, ema_b}, and thus the HB is usually 
carefully avoided in the interpretation of either the age or the metallicity properties of a resolved stellar population.

This is particularly unfortunate for the study of the ancient ($>$10~Gyr old)
stellar populations in nearby galaxies.   
These populations are the fossil record of the early stages of galaxy evolution, 
and provide crucial information about the epoch of formation of cosmic structures. 
The standard age diagnostics for these stars are to be found at faint magnitudes in a CMD, 
compressed into the oldest main sequence (MS) turn off region. This region of the CMD, apart from being small 
(in terms of magnitude and colour extension) and thus very sensitive to photometric errors, may also contain 
overlapping younger populations. Also, the age-metallicity degeneracy, 
although much better behaved than on the RGB, may still create problems
in determining a unique solution for the ages and metallicities of the
ancient stars.

Stellar evolution suggests that, at least in principle, 
the detailed properties of stars older than $\sim$10~Gyr could be recovered by 
analyzing the much more luminous and extended HB, if only its morphology 
could be sorted out in terms useful for understanding the relation to the star 
formation history and metallicity evolution of the galaxy.

The purpose of this work is to see whether it is possible to
accurately model the resolved HB of the Sculptor dwarf spheroidal
galaxy taking into account its past star formation history (SFH -- Here we denote with SFH the star formation rate 
as a function of age and metal content). 
Sculptor has an exceptionally detailed SFH determination -- derived without considering the HB -- obtained by combining CMD analysis with detailed 
spectroscopic metallicities along the RGB \citep{deboer12}. 
We also use the well measured RR~Lyrae properties and the optical CMD of non variable objects, as constraints 
to determine the best synthetic HB populations that match the observed CMD, using BaSTI stellar 
models \citet{basti, basti2}.
The only free parameter in our analysis is the total amount of mass lost along the RGB, given that 
age and metal abundance distributions are fixed by the SFH. 
To the best of our knowledge this is the first time that a complete and detailed modelling of the HB of a resolved galaxy 
has been performed; our investigation is therefore a first step to assess the potential of 
synthetic HB modelling to add further constraints on a galaxy SFH, an issue particularly important when studying galaxies so distant 
that only RGB and HB stars can be resolved. 

%We will test whether the SFH derived from the MS and RGB reproduces the {\sl correct} number of HB stars, and also whether 
%by varying just the RGB total mass loss it is possible to reproduce the global stellar distribution (in terms of colour and magnitudes) 
%along the whole observed HB. 
Any mismatch between the observed and predicted magnitude distribution along the HB may be a signature of the presence of 
enhanced-He populations -- the counterpart of second generation stars in Galactic globulars -- 
not included in the SFH determinations. This will provide an important additional piece of information 
regarding the comparison of photometric and chemical properties of dwarf galaxies and globular clusters. To date, 
comparisons of chemical abundance patterns between Galactic globulars and dwarf galaxies reveal that these latter lack  
the abundance anticorrelations (e.g. O-Na) typical of second generation stars in individual globular clusters 
\citep[see, i.e.,][]{gei}, that are associated to varying degrees of He enhancement \citep[see, i.e.,][]{pasquini}. 
By {\sl enhanced-He} populations we mean stars born with initial He mass fractions (Y) larger than the cosmological  
He abundance Y=0.246 \citep[see][for a recent reevaluation]{coc}.
The BaSTI models employ initial Y values that scale with 
the metal mass fraction Z as dY/dZ$\sim$1.4, derived by 
considering a cosmological Y$=$0.245 and the initial solar Z and Y as derived from a calibration of the standard solar 
model \citep[see, e.g.][]{basti}. In the metallicity regime of Sculptor the resulting Y values are practically constant -- 
ranging between 0.245 and 0.248 -- and essentially equal to the cosmological He.
It is important to notice that the determination of the SFH -- see next section -- 
employed the same dY/dZ scaling, hence approximately the cosmological He, across 
the whole Z range covered by the galaxy population.

The RGB mass loss law derived from the HB modelling -- eventually a function of initial metallicity and/or age -- 
will also provide baseline values to be compared with analogous determinations on globular clusters, and to  
be tested on the HB of other dwarf galaxies.
As a byproduct of our investigation we will compare in the appendix the observed difference between RGB bump and HB magnitudes, with the theoretical 
prediction based on the galaxy SFH, and verify whether a discrepancy exists, as found in Galactic globulars 
\citep[see, e.g.,][]{dicecco}.

The paper is structured as follows. Section~2 describes briefly the data and the SFH used in this investigation, whilst 
the next two sections present our synthetic HB analysis and results, followed by our conclusions.
 
\section{Data}
\label{data}

The Sculptor dwarf galaxy is well known to contain a predominately old stellar
population \citep[see, e.g.,][]{dacosta}, so the contamination of populations $<$10~Gyr old along the 
MS and overlapping the HB is minimal.
Early work on the Sculptor HB by \citet{bumpSc} was mainly qualitative
and made use of simple ZAHB fitting, to assess the galaxy metallicity
distribution. It was later confirmed that the red and blue HBs in Sculptor exhibit
differences in spatial distribution that correspond to differences in
ages and metallicities within the oldest population \citep{Tolstoy04, deboer11}. These previous works simply highlighted 
that the HB is complex, and age and metallicity spreads play a role in creating the complexity.

A healthy HB extended in colour from the red to the blue  
indicates that a significant number of RR~Lyrae variable stars must populate the instability strip, as is observed. 
The most complete survey to date comes from \citet{kal:95}, where 226 RR~Lyrae stars were 
identified and their light curves classified. Their properties are consistent with a spread in metallicities, and [Fe/H]$< -1.7$. This was 
confirmed by \citet{clementini} low resolution spectroscopy of 107 variables, which showed 
the metallicity to peak at  
[Fe/H]$\sim -1.8$, with a range covering $-2.40 <$ [Fe/H] $< -0.8$.
 
For the accurate modelling of the resolved HB of the Sculptor dwarf spheroidal, we make use of a deep optical B,V CMD 
presented in \citet{deboer11}. In the V-(B-V) CMD the HB 
is the most horizontal, hence most sensitive to potential He variations. The observed HB extends to (B-V)$\sim$0 in the blue, 
and even at these colours a change of Y (at fixed Z) affects both the luminosity of the zero age HB (ZAHB) and 
the luminosity of the end of the HB phase, 
thus enabling to test the presence of enhanced-He stars also along the bluest part of the galaxy HB. 

The photometry was obtained using the CTIO 4-m MOSAIC II camera, and 
carefully calibrated using observations of 
Landolt standard fields~\citep{Landolt07, Landolt92}. This resulted in an accurate photometric catalogue, covering a region 
ranging from the Sculptor centre out to an elliptical radius r$_{ell}$$\le$1~deg. 
The HB of Sculptor can be seen not have an extended blue tail in the optical, and 
the hottest ${\rm T_{eff}}$ is lower than the observed limit ($\sim$ 12000~K) for the onset of radiative levitation. 

The synthetic HB calculations also make use of the detailed SFH of Sculptor determined by \citet{deboer12}. 
This SFH is obtained using the 
same optical photometry presented in \citet{deboer11}, combined with detailed spectroscopic metallicities along the RGB, not using the  
observed HB stars.  
The distance modulus ${\rm (m-M)_V}$=19.72 and reddening E(B-V)=0.018 used in the SFH determination are also employed  
in the HB modelling. This SFH determination uses 
the DSEP stellar evolution models \citep{dsep} -- that employ the same value as BaSTI for the 
cosmological He, and a very similar dY/dZ$\sim$1.5, that produces essentially the same 
initial Y for Sculptor metallicity range --  and there is good agreement between the 
Sculptor SFH determined with DSEP or BaSTI models, as 
also shown by \citep{deboer12}.

The SFH is provided for 5 annuli, extending from the centre out to a radius ${\rm r_{ell}}$=1~deg.  The region of interest for our simulation comprises 
the innermost two annuli, with ${\rm r_{ell}} \le$0.183~deg, that include the survey of RR Lyrae variable stars by \citet{kal:95}. 
The adopted star formation rate  within ${\rm r_{ell}} \le$0.183~deg is displayed in Figs.~2 and 13 of \citet{deboer12}; 
it is provided in terms of solar masses per year in 1~Gyr age bins, between 5 and 14~Gyr. 
Broadly speaking, the star formation rate peaks in the oldest age bin, and then declines slowly with time; it reaches half the peak value 
between 9 and 10~Gyr ago, and decreases down to zero between 5 and 6~Gyr ago. The low star formation rate means that 
the age bins between 5 and 8~Gyr 
have a negligible impact on the resulting synthetic HB. 

For each age bin the SFH is further subdivided into 8 [Fe/H] bins 0.2~dex wide, covering the range 
between [Fe/H]=$-$2.5 and $-$0.9. The mean value of [Fe/H] increases slowly with the mean age of the population. 
%They are 0.2~dex wide but for the most metal rich one, that ranges between [Fe/H]=$-$1.1 and $-$0.9.
A range of [$\alpha$/Fe] values (0.2 or 0.3~dex wide) is associated to each [Fe/H] bins 
\citep[see Fig.~2 of][]{deboer12}. The values of [$\alpha$/Fe] 
decrease slowly with increasing [Fe/H] and are generally positive, 
but for the more metal rich bins, where [$\alpha$/Fe] also reaches negative values down to $-$0.2~dex for the higher [Fe/H] bin. 

The properties of RR Lyrae stars are taken into account using \citet{kal:95} survey in the centre of Sculptor. 
The survey employed many short expsoure images covering a wide range in time to identify the RR Lyrae stars in the central region of Sculptor and determine the 
light curve profiles. Their sample can be considered complete in terms of photometric and temporal sampling of the variable objects on the HB. Intensity-weighted 
magnitudes are provided in the V band, averaged over the pulsation period of each pulsator. Furthermore, to prevent contamination of non-variable HB features by 
the presence of variable stars at random phase, we have identified and removed the confirmed RR Lyrae variable stars from \citet{deboer11} data -- 
being taken at random phase, these objects were scattered to the blue, red and within the instability strip --  
based on spatial position.
Due to the square field of 
view of the \citet{kal:95} fields,  compared to the elliptical region considered here, the area covered by the SFH is $\sim$30\% larger than the 
area for which RR Lyrae measurements are available. 
Therefore, we will apply a 30\% reduction of the number of synthetic stars in the instability strip, in comparisons 
with the observed RR Lyrae populations, to account for the different size of the area covered by \citet{kal:95} observations. This also 
implies that there could be a relatively low number of undetected RR Lyrae variables taken at random phase (about 70 at most) 
scattered along and around the HB, that however do not affect appreciably the number distributions used for our analysis.  
As suggested by the referee, we have performed the full analysis described in the next sections also for the objects with ${\rm r_{ell}} \le$0.116~deg, e.g.,  
within the innermost region of \citet{deboer12} SFH determinations. This reduces by a factor $\approx$2 the observed number of HB objects 
used in the analysis. In this case \citet{kal:95} field covers the whole region, implying that the RR Lyrae sample is essentially complete 
and also all variables can be removed 
from \citet{deboer11} photometry. The results we obtain are completely consistent 
with what found for the case of ${\rm r_{ell}} \le$0.183~deg, that will be detailed in the next sections.

\section{Methods}

We describe in this section the methods employed in our theoretical modelling of Sculptor HB.

\subsection{Synthetic CMD generation}

In a complex old stellar population, the morphology of the HB will depend on both the chosen RGB mass loss and the 
input SFH. For this reason we have completely rewritten the BaSTI synthetic CMD generator \citep[SYNTHETIC MAN --][]{basti} 
to account for both a generic SFH and an arbitrary RGB total mass loss.
The code now produces a full synthetic CMD that includes also -- for old populations --  a HB of arbitrary morphology. 

Briefly, the code reads first the full grid of models -- tracks from the MS to the tip of the RGB  
and HB tracks -- from the BaSTI database for varying Z and a 
scaled solar metal distribution \citep{basti}, and the SFH of the population --  in our case the SFH by \citet{deboer12}. 
For each age and [Fe/H] bin the synthetic CMD generator first draws randomly a stellar mass $m_i$, within the range 0.1 and 120  
${\rm M_\odot}$ \citep[for consistency with][]{deboer12}, according to a \citet{kroupa} initial mass function. 
A value of the age $t_i$ and 
${\rm [Fe/H]_i}$ are then drawn randomly according to uniform probability distributions 
\citep[in agreement with the derivation of the SFH by][]{deboer12}, within the ranges associated to the selected age and [Fe/H] bin. 
In addition, a value of ${\rm [\alpha/Fe]_i}$ is also drawn randomly (uniform probability distribution) within 
the range prescribed by the SFH for the selected age and [Fe/H] bin.

If the model corresponding to $m_i$ and the selected 
metal composition has a lifetime at the first thermal pulse or at carbon ignition  
-- the relevant age range for Sculptor SFH is fully covered by the BaSTI models -- 
shorter than $t_i$, the star won't appear in  
the synthetic CMD, and a new set of $m_i$, ${\rm [Fe/H]_i}$, ${\rm [\alpha/Fe]_i}$ and $t_i$ is drawn. 
We remark that for a given ${\rm [Fe/H]_i}$ and ${\rm [\alpha/Fe]_i}$ 
we consider scaled solar models with total metallicity ${\rm [M/H]\sim[Fe/H]+log(0.638 \ 10^{[\alpha/Fe]}+0.362)}$, 
that closely mimic models with the same [M/H] and varying [$\alpha$/Fe] in 
the low mass, low metallicity regime covered by the stars currently evolving in this galaxy \citep{scs:93}.

If the lifetime of the model with mass $m_i$ 
is larger than $t_i$, and $t_i$ is smaller than the model age at He-ignition, interpolation 
in age and Z amongst the BaSTI tracks will determine the CMD location of this synthetic star. 
If $t_i$ is larger than the lifetime at He-ignition, but smaller than the lifetime at the first thermal pulse,  
the code subtracts from $m_i$ a specified value of the total mass loss along the RGB $\Delta M_{RGB}$, 
to provide a value for $m_{i, HB}$, e.g., the HB mass of the synthetic star with initial mass $m_i$. 
Interpolations in mass, Z and age amongst the BaSTI HB tracks provide the CMD location of the synthetic star.

Finally, the synthetic magnitudes determined with the described procedures, are perturbed 
by a mean Gaussian photometric error to approximate the observational 
error of \citet{deboer11} data. 
For the magnitude range of the HB stars, the typical 1$\sigma$ photometric uncertainty of non-variable stars is 
$\sim$0.007 mag in B and V.

The values of $m_i$ are added, until the total mass of stars formed in each SFH bin satisfies the derived star formation rates.
We have often arbitrarily multiplied the star formation rates in each bin by constant factors, 
to have a larger population of HB stars and minimize the Poisson noise in the simulations.

It is clear from this brief sketch of the synthetic CMD calculation, that 
the only free parameter entering our analysis is $\Delta M_{RGB}$.   

\begin{figure}
\centering
\includegraphics[width=\columnwidth]{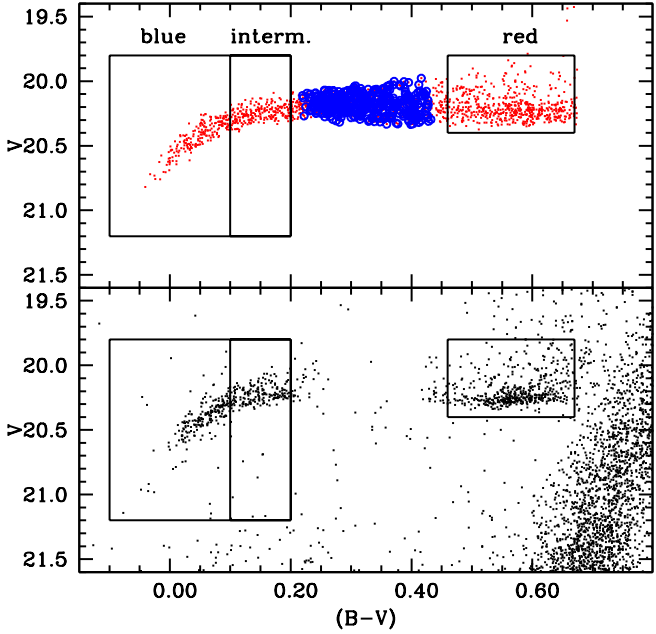}
\caption{Synthetic (top) vs observed (bottom) CMD for the region in Sculptor dwarf spheoridal within ${\rm r_{ell} < 0.183}$ deg. 
The synthetic CMD has a number of HB stars close to the observed 
number (see text for details of the simulation). The three boxes (red, intermediate and blue)  
mark the three areas where star counts are compared with the simulations.
The empty (blue) circles with dot represent the RR Lyrae stars found in the simulations.
The observed RR Lyrae stars (as they are taken at random phase) are excluded from the bottom panel CMD.
There are some non-pulsating stars in both synthetic and observed samples, 
located between the {\sl red} and {\sl intermediate} blue boxes (see text for details).}
\label{CMD}
\end{figure}

\subsection{Modelling RR Lyrae stars}

For the RR Lyrae stars observed by \citet{kal:95} only the 
mean V magnitudes of individual pulsators are available. Therefore, we had to employ theoretical determinations of the instability strip (IS)  
colour boundaries in our synthetic CMDs, together with the pulsational equation \citep{dcr:04} that provides the fundamental period (P) 
of a synthetic object as a function 
of its mass, metallicity, bolometric luminosity and effective temperature. First overtone 
periods (${\rm P_{FO}}$) are related to the fundamental ones by the relation  P=${\rm P_{FO}}$+0.13 \citep{dcr:04}.
The boundaries of the fundamental (F) and first overtone (FO) strips were taken from 
\citet{dcr:04}, with small adjustments -- within the quoted theoretical uncertainties -- to account for the empirical constraints given by the 
period distribution determined by \citet{kal:95}, that poses strong 
constraints on the width of the IS. As an example, if the theoretical red boundary is too red, 
the synthetic objects reach too long periods. 
We have therefore {\sl adjusted} the boundaries of the F and FO regions 
in order to match as well as possible the period range covered by the observed F and FO pulsators.
When using the {\sl standard} IS boundaries provided by \citet{dcr:04} from calculations with 
mixing length {\sl ml}=1.5${\rm H_p}$, the synthetic objects reached too long periods. This constraint  
forced us to consider a larger value of {\sl ml} in the pulsational results, and that shifts the red edge of the FF region 
to hotter temperatures, and lowers the upper boundary of the F periods. Thus we have 
employed the derivatives provided by \citet{dcr:04}, to determine the IS edges at varying {\sl ml}. 
However, an increased {\sl ml} tends to move the blue edge of the FO region to the red, narrowing down 
too much the portion of the IS populated by FO pulsators. We therefore further adjusted 
the F and FO boundaries at fixed {\sl ml}, within their nominal uncertainties 
of the order of $\pm$50-100~K \citep{mm:03}. 

The final V magnitudes of the synthetic stars that lie within the IS will be compared to 
the RR Lyrae observations (their mean V magnitudes) by   
\citet{kal:95} paper. The photometric uncertainty on the V magnitude of the observed variables is greater then the photometric 
uncertainty of \citet{deboer11} data, due to the shorter exposure times employed by \citet{kal:95}. 
In the comparison with the RR Lyrae V distribution, to match the observational 
conditions of these data, the V magnitudes of the synthetic stars that lie within the IS have been perturbed by a Gaussian photometric 
error with $\sigma$=0.05 mag, as determined from Fig.~1 of \citet{kal:95}.

\subsection{Comparing models and observations}

The bottom panel of Fig.~\ref{CMD} displays the observed HB, divided into three regions, delimited by rectangular boxes, 
that we denote as {\sl red}, {\sl intermediate} and {\sl blue}. These boxes contain, respectively, 457, 194, and 401 stars.
The bluer boundary of the {\sl red} box and the red boundary 
of the {\sl intermediate} box have been drawn at the approximate colours of the boundaries of the instability strips in 
Galactic globular cluster CMDs. The region between the {\sl red} and {\sl intermediate} box contains mainly 
RR Lyrae variables, but also a few non variable 
stars displayed in the CMD\footnote{In case of our simulations for the innermost region with ${\rm r_{ell}} \le$0.116~deg, that is  
fully covered by 
\citet{kal:95} RR Lyrae observations, we find also some stars in the CMD region between {\sl red} and {\sl intermediate} boxes. This strenghtens  
the case for the small number of objects with ${\rm r_{ell}} \le$0.183~deg lying in the same CMD region, to be mostly non-variable stars, rather 
then -- another viable possibility -- RR Lyrae objects taken at random phase and not covered by \citet{kal:95} field of view}.
We considered a synthetic HB model for a given $\Delta M_{RGB}$ choice to be satisfactory match to the observations when: 

\begin{enumerate}

\item{the relative number of stars in the three boxes and within 
the region containing the RR Lyrae IS is reproduced within the Poisson uncertainty;}

\item{the observed mean V-magnitude of the non variable stars in each of the three boxes is matched within 0.01~mag;}

\item{the total (B-V) extension of the observed HB is well reproduced.}

\end{enumerate}

When these conditions were satisfied, we also checked that there was general agreement 
with the overall shapes of the histograms of observed star counts as a function of both V and (B-V) for the non variable stars, 
and with the mean level of the RR Lyrae brightness (averaged over a pulsational cycle).  

Two points must be mentioned regarding these additional constraints. 

First, we could in principle have enforced the constraint of perfect statistical agreement between the theoretical and observed 
star counts as a function of both V magnitude and colour for the non variable objects. 
However, a perfect fit depends on a precise knowledge of the functional dependence of $\Delta M_{RGB}$ 
(and eventually the initial Y) on one or more stellar parameters. Given the current lack of solid 
theoretical and empirical guidance, this 
dependence may be extremely complicated, discontinuous or involving additional physical processes not included in stellar models. 
We have therefore used simple parametrizations that, as we will see in the next section, produce synthetic CMDs that are able 
to satisfy our three main criteria listed above. In this way we put strong constraints on $\Delta M_{RGB}$ 
and Y.

Regarding the RR Lyrae mean magnitudes, they were determined from independent observational data. The calibrated V magnitudes  
have a maximum zero point systematic uncertainty of $\sim$0.03~mag, according to \citet{kal:95}, whilst this  
uncertainty is much smaller for the non variable star photometry \citep{deboer11}.
Given a possible small zero-point mismatch between the two photometries, we have simply checked  
that, when the main criteria were satisfied, the mean level of the synthetic RR Lyrae sample 
matched the observed one within the uncertainty on the relative zero-point of the two photometries.
As a final, albeit weaker consistency check, we also compared the predicted and observed period distributions, 
and the predicted RR Lyrae [Fe/H] distribution, with the spectroscopic observations of \citet{clementini}.

\section{Results}

We have calculated several synthetic CMDs, each one with typically much larger numbers of stars than observed -- 
by scaling appropriately the SFH  --  
to minimize the Poisson error on the synthetic star counts. For each test we fixed $\Delta M_{RGB}$, and checked  
the agreement between synthetic and observed HB morphology. 
As a zero order approximation, we tried with a standard 
Gaussian $\Delta M_{RGB}$ and mean value (and 1$\sigma$ spread) independent of [M/H] and age. This mass loss prescription  
did not match the observations, for any choices of the mean value and $\sigma$ spread around the mean. 
Also the use of a uniform probability with varying mean values and $\sigma$ spread, both independent of [M/H] and age did not 
help. 
The analysis of these failed attempts demonstrated that $\Delta M_{RGB}$ must be made dependent on [M/H], and also that 
only a small dispersion around these metallicity dependent mean values is consistent with the observations.

\begin{figure}
\centering
\includegraphics[width=\columnwidth]{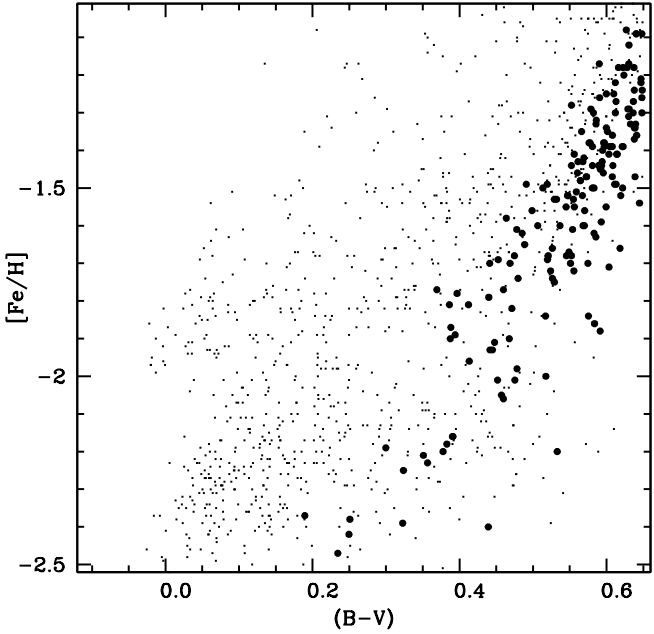}
\caption{[Fe/H] as a function of (B-V) for the best fit synthetic HB sample. Dots (filled circles) denote 
stars with age equal or larger (smaller) than 10~Gyr.  (see text for details).}
\label{fehBV}
\end{figure}

The simulations that matched the observed HB according to the criteria described in the previous section 
had a Gaussian distribution of $\Delta M_{RGB}$ with mean values:

\begin{enumerate}

\item{$<\Delta M_{RGB}>$=0.10${\rm M_{\odot}}$, for [M/H] $< -$1.8}  

\item{$<\Delta M_{RGB}>$=0.14${\rm M_{\odot}}$ , for $-1.8\le$ [M/H]$\le -$1.6}

\item{$<\Delta M_{RGB}>$=0.14 - 0.15${\rm M_{\odot}}$ , for $-1.6 <$ [M/H]$\le -$1.4}

\item{$<\Delta M_{RGB}>$=0.14 - 0.16${\rm M_{\odot}}$ , for $-1.4 <$ [M/H]$\le -$1.3}

\item{$<\Delta M_{RGB}>$=0.16${\rm M_{\odot}}$, for [M/H]$> -$1.3}   

\end{enumerate}

and a very small dispersions, also metallicity independent, $\sigma$=0.005 $M_{\odot}$. 
We determined $<\Delta M_{RGB}>$ taking into account the error bars on the best-fit star formation rates -- as provided 
by \citet{deboer12} -- 
hence the range of values for $-1.8\le$ [M/H]$\le -$1.3. In the other metallicity 
ranges the uncertainty on the star formation rates causes $<\Delta M_{RGB}>$ variations below 
0.01${\rm M_{\odot}}$. All discussions and figures that follow display results obtained from simulations 
with the best-fit value of the star formation rate.  

In place of step functions for $<\Delta M_{RGB}>$ vs [M/H] 
we tried also linear or quadratic analytical expressions as a function of only [M/H] or both [M/H] and age,  
but the match to the observations got generally worse.

Although the Reimers law \citep{reimers} -- still widely employed to calculate RGB mass loss rates -- 
predicts slightly increasing values of $<\Delta M_{RGB}>$ at increasing metallicity,  
when the free parameter $\eta$ that enters Reimers formula is kept fixed, 
the variations necessary to model Sculptor HB are larger. To this purpose, using the BaSTI models,  
we checked how our derived $<\Delta M_{RGB}>$ values can be transposed into values of the parameter $\eta$.
For [M/H] $< -$1.8  $<\Delta M_{RGB}>$ is slightly higher than what predicted by $\eta$=0.2 (that provides 
integrated RGB mass loss values between 0.06 and 0.07$M_{\odot}$, depending on the age), whilst 
for [M/H]$> -$1.8  $<\Delta M_{RGB}>$  is close to what is predicted by $\eta$=0.4.

With this mass loss calibration the mean V magnitudes in the {\sl red}, {\sl intermediate} and {\sl blue} boxes 
are equal to, respectively, 
20.20, 20.25 and 20.33, identical within 0.01~mag to the observed values. The mean brightness of the RR Lyrae IS 
is within 0.03~mag (the simulation being fainter) of the observed value, equal to 20.14~mag.
The top panel of Fig.~\ref{CMD} displays one realization of the HB, without scaling the galaxy SFH.  
The total absolute number of HB objects and the absolute number counts (not just the relative ones) 
in the three boxes and the region of the IS turn out to be equal -- within the Poisson errors on the star counts -- to 
the observations (again, considering the scaling factor for the limited area of the observed 
RR Lyrae population). This is an extremely important independent check of the derived star formation rates.

Figure~\ref{fehBV} displays [Fe/H] (more directly linked to observations than [M/H]) 
as a function of (B-V), for the HB stars in the simulation of Fig.~\ref{CMD}. 
There is a general trend of decreasing [Fe/H] with decreasing colour, in agreement with the early analysis by \citet{bumpSc} 
that made use of ZAHB fitting to the observed HB. However, the [Fe/H] dispersion at a given (B-V) is very large.
If we split the synthetic sample into objects older than 10~Gyr and younger ones, we obtain 
${\rm [Fe/H]=1.46 \ (B-V) - 2.29 }$ for the old sample, and a much steeper dependence ${\rm [Fe/H]=2.80 \ (B-V) - 3.10 }$ 
for the young objects, as is also clear from Fig.~\ref{fehBV}. 
The dispersion of [Fe/H] around these mean relationships is equal to 0.31~dex for the old sample, 
and 0.18~dex for the young one.

\begin{figure}
\centering
\includegraphics[width=\columnwidth]{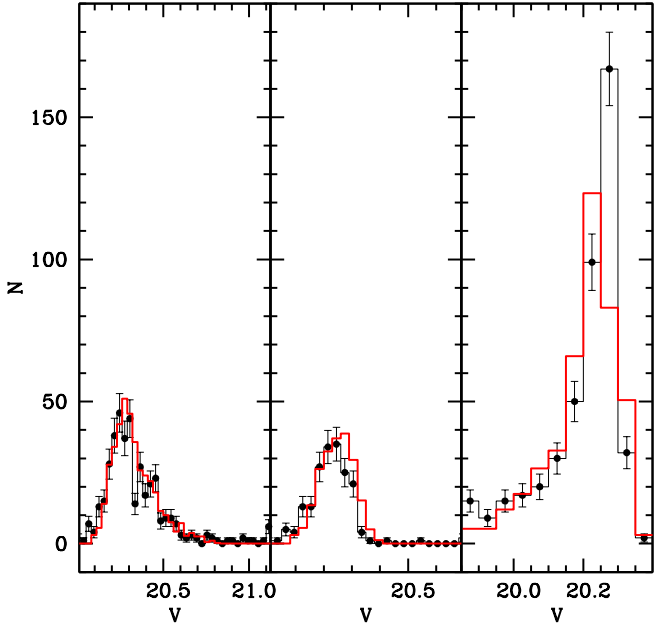}
\caption{Observed (solid black line and filled circles) vs synthetic star counts (red line) as a function of the V magnitude 
in the (from left to right) {\sl blue}, {\sl intermediate} and {\sl red} boxes, respectively (bin size of 0.05~mag 
for stars in the {\sl red} box and 
0.03~mag for the other two samples). Poisson errors on the observed star counts are 
also displayed.}
\label{Vhist}
\end{figure}

\begin{figure}
\centering
\includegraphics[width=\columnwidth]{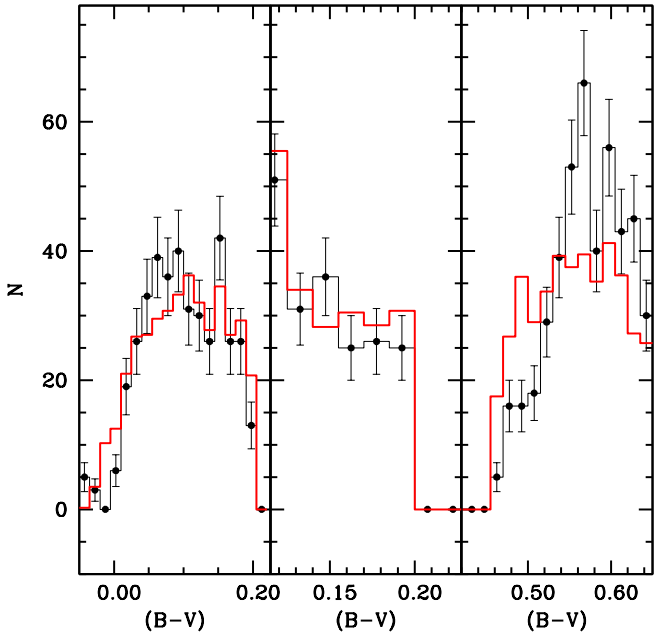}
\caption{As Fig.~\ref{Vhist}, but for the (B-V) colour (bin size of 0.015~mag).}
\label{BVhist}
\end{figure}

Figures~\ref{Vhist} and \ref{BVhist} compare observed and synthetic star counts as a function of V and (B-V), respectively.
Here we used simulations with a much larger number of objects than observed, and rescaled the star counts 
appropriately, to compare with observations.

The overall shape of the theoretical histograms reproduces very closely the observed one.  
The comparison is worse for the star counts as a function of colour in the {\sl red} box.
It is remarkable that the magnitude distributions, 
very sensitive to the initial chemical composition, are well matched by the simulations, not just the mean values of V in the three 
boxes. There is no significant excess of stars brighter than what is predicted by the simulations, hence there is no clear evidence  
of He-rich stars.

\begin{figure}
\centering
\includegraphics[width=\columnwidth]{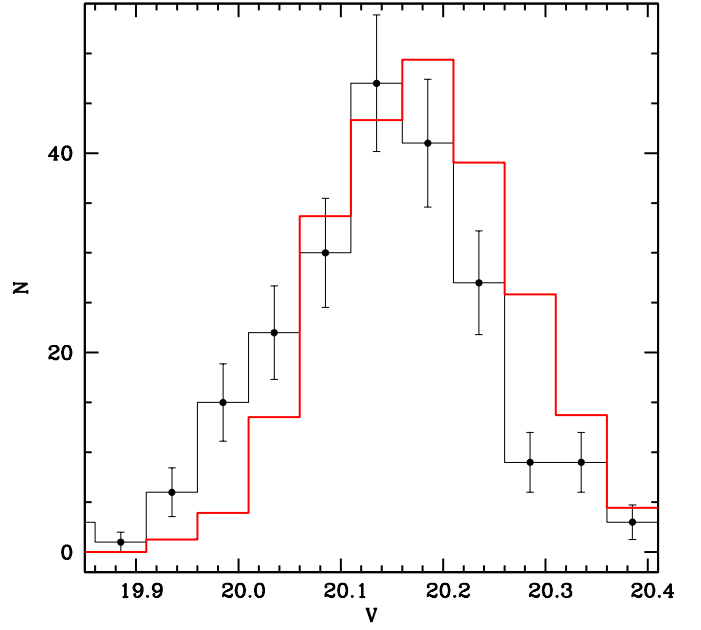}
\caption{As Fig.~\ref{Vhist}, but for the RR Lyrae population (bin size of 0.04~mag).}
\label{RRV}
\end{figure}

We now consider comparisons with the RR Lyrae IS, for which not only mean V magnitudes, but also 
[Fe/H] estimates from \citet{clementini} are 
available, to provide 
an additional consistency check of the HB simulations. These authors determined [Fe/H] for about half of the 
variables observed by \citet{kal:95}, using a revised version of the ${\rm \Delta S}$ method \citep{ds}, calibrated 
on both the \citet{zw:84} and \citet{cg:97} (hereafter ZW and CG, respectively) 
globular cluster [Fe/H] scales. Typical errors on individual [Fe/H] estimates are 
equal to $\sim$0.15~dex.
  
Figure~\ref{RRV} compares observed and synthetic RR Lyrae star counts as a function of V. The overall shapes 
of the histograms are very similar, with just a slight offset of the synthetic sample towards fainter magnitudes, as 
indicated by the mean brightness, fainter by 0.03~mag. This offset is equal to the estimated maximum zero point error on the 
calibration of the observations \citep{kal:95}.

\begin{figure}
\centering
\includegraphics[width=\columnwidth]{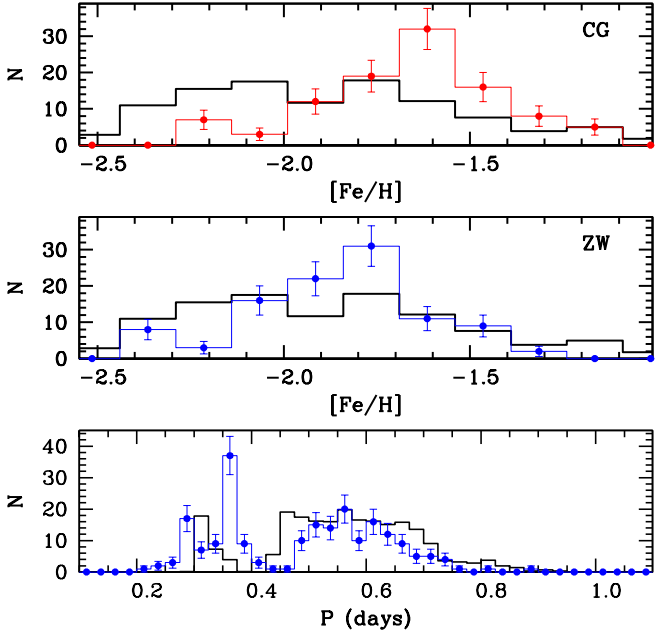}
\caption{Comparison with observations (displayed as histograms and filled circles with error bars) of star counts in the IS, as a function of P 
(bin size of 0.025 days) and [Fe/H] (bin size of 0.15~dex). Results for both 
the \citet{zw:84} and \cite{cg:97} [Fe/H] scales are displayed in the middle and top panel, respectively.}
\label{RRPFEH}
\end{figure}

Period \citep[data from][]{kal:95} 
and [Fe/H] distributions for the sample of variables are displayed in Fig.~\ref{RRPFEH}. The theoretical star counts have been rescaled to 
account for the smaller number of objects with [Fe/H] determinations, compared to the photometric sample of RR Lyare stars. 
The fit to the period distribution is not perfect even after tuning -- within the constraints  
imposed by pulsational models -- the boundaries of the IS to match as well as possible 
the range covered by the observations. Only an {\sl ad hoc} tuning of the F and FO boundaries would provide a better match.
As noticed by \citet{RRproc}, the IS boundaries would be probably better determined with a lower {\sl ml} 
for FO pulsators compared, to F ones. This would certainly improve the fit of the P histogram, that displays 
a substantial discrepancy for the ratio of FO to F pulsators, such that it is too low in the simulation. In this respect one also has to consider 
that we could not take into account the objects that are in the so called OR zone of the IS, where stars pulsate 
FO or F depending on where they evolved from (FO or F region). This adds an additional uncertainty 
when trying to match precisely the observed period distribution.

Regarding the metallicity distribution, we display the results for both the ZW and CG [Fe/H] scales. 
On the whole, the synthetic sample matches the observed values on the ZW scale fairly well. The average [Fe/H] 
of the synthetic sample is ${\rm <[Fe/H]>}=-$1.88, very close to the observed one on 
the ZW scale, equal to ${\rm <[Fe/H]_{ZW}>}=-$1.84. 
For comparison, the observed mean value on the CG scale is ${\rm <[Fe/H]_{CG}>}=-$1.65

All these results are based on CMD simulations of the central regions of the galaxy, 
covered by the RR Lyrae photometry. It is impossible to constrain 
completely the synthetic HB models without information about the variable star population within the IS.
We can however check whether our mass loss calibration based on the inner regions, is able to match the mean magnitudes and 
observed number of HB stars in the {\sl blue}, {\sl intermediate} and {\sl red} boxes of Fig.~\ref{CMD}, when the 
whole observed area from the centre out to an elliptical radius ${\rm r_{ell}}=$1~deg is considered\footnote{We cannot 
eliminate from the observed CMD the RR Lyrae stars taken at random phase, located at ${\rm r_{ell}}>$0.183~deg, because we cannot identify them}.
To this purpose we calculated synthetic CMDs as described, considering the appropriate SFH for the whole observed area, using the same 
integrated RGB mass loss prescription of our best fit to the HB of the inner region.
As a result, we can match within 0.01~mag the observed mean V magnitudes of stars in the three boxes, 
and the empirical star counts (within the associated Poisson error) in these three regions of the CMD.

\section{Conclusions}

We have performed the first detailed simulation of the HB of a resolved dwarf galaxy, taking consistently into account 
the SFH determined from MS and RGB photometric and spectroscopic observations, and using 
synthetic HB techniques usually applied to study the HB in globular clusters. 
The number of HB stars predicted by our simulations is consistent with observations, within the Poisson error on the star counts.
The colour and V-magnitude distribution of all non-variable HB stars in Sculptor is matched well by the synthetic model, for simple choices of the 
integrated RGB mass loss. This latter needs to be metallicity dependent -- with a stronger dependence 
than predicted by the Reimers law at fixed $\eta$ -- to satisfy the observational constraints, and also must have  
a very small dispersion at fixed metallicity. 
The magnitude, metallicity (on the ZW scale) and period distribution of the RR Lyrae stars are also satisfactorily reproduced, when taking 
into account the current uncertainties on the IS boundaries.

There is no indication of enhanced-He subpopulations along the HB from the V-magnitude distribution of the non-variable and variable stars 
-- within the uncertainty on the relative photometric zero-point. The metallicity range covered by the SFH, 
as constrained by spectroscopy of RGB stars, plus a simple RGB mass loss law, enable to cover both the full magnitude and colour range of HB stars. 
There is no excess of bright objects to be matched with enhanced-He populations. The agreement of the synthetic model with 
observations dictates that any enhancement of He -- if present -- has to be lower than $\Delta$Y=0.01.
The good agreement of the synthetic sample with the [Fe/H] distribution of the RR Lyrae stars (on the ZW scale) lend 
additional support to the results of our simulations. 

The lack of signatures of enhanced-He stars along the HB is consistent with the lack of the O-Na anticorrelation 
observed in Sculptor and other dwarf galaxies, and confirms the intrinsic difference between Local Group dwarf galaxies and globular 
cluster populations.

Regarding the RGB integrated mass loss $<\Delta M_{RGB}>$, our simulations suggest a very simple prescription for the case of Sculptor, 
with $<\Delta M_{RGB}>$ slowly increasing with increasing [M/H]. We also find an extremely small spread around these mean values. 
It is natural to try and compare $<\Delta M_{RGB}>$ determined from our analysis, with 
similar estimates in globular clusters to assess whether, at least in case of FG stars, 
the RGB mass loss is approximately the same as in this galaxy. We compare here with the results by 
\citet{ema, ema_b}, who used synthetic HB models (using BaSTI models) to determine initial Y and mass distribution 
of HB stars in a few globular clusters. 
For NGC~2808, a cluster with [Fe/H]$\sim -$1.2 ([M/H]$\sim -$1.0, when considering a standard 
value [$\alpha$/Fe]=0.3-0.4), 
$<\Delta M_{RGB}>$=0.15${\rm M_{\odot}}$ for FG stars, that compares well with the value obtained for 
Sculptor at this metallicity.
In case of M3 and M13, both with [Fe/H]$\sim -$1.6, the HB analysis provides 
$<\Delta M_{RGB}>$=0.12${\rm M_{\odot}}$ and 0.21${\rm M_{\odot}}$, respectively. 
Whereas the value for M3 is reasonably close to what we derive for Sculptor at the appropriate 
metallicity ($<\Delta M_{RGB}>\sim$0.14${\rm M_{\odot}}$), the result for M13 is obviously very discrepant.
On the other hand, even when $<\Delta M_{RGB}>$ values are similar to Sculptor, the dispersion around these mean values 
is usually larger in the globular clusters. 
Before drawing any strong conclusions about the similarity (or lack) of the integrated RGB mass loss 
in globular cluster FG stars and dwarf galaxies, it is 
clearly necessary to extend analyses like ours and \citet{ema, ema_b} to much larger samples of objects.

In summary, our results show that in case of Sculptor a simple
mass loss law is able to explain the observed detailed HB morphology. The next obvious step is to verify whether synthetic HB models with
the same mass loss law can reproduce the HB of other resolved dwarf galaxies, with well established and diverse SFHs.
If this is the case, the combined analysis of the RGB (strongly affected by the initial metallicity) and
the HB in distant galaxies where only these phases can be resolved, will be able
to provide constraints on the age and metallicity distributions of their oldest populations. 

Finally, as a byproduct of our simulations, we have been able to compare the observed magnitude of the RGB bump 
with theory. As detailed in the appendix, 
we find a discrepancy between observed and theoretical RGB bump, consistent  
with results for Galactic globulars, that point towards a too bright bump in stellar models, at least 
for intermediate metallicity and metal poor clusters.
This is at odds with recent results by \citet{monelli10} for Sculptor, that were however based on older photometric data.
We are able to explain the difference with our results for this galaxy, but 
the analysis of the RGB bump brightness in dwarf galaxies clearly deserves further detailed investigations.

\begin{acknowledgements}
We thank the anonymous referee for several comments that improved the presentation of our results.
MS wishes to dedicate this paper to the memory of his father, who sadly passed away a few months ago.
He also thanks the Kapteyn Astronomical Institute for their hospitality and support of a visit, during which an important 
point of this analysis was clarified.       
SC is grateful for financial support
from PRIN-INAF 2011 "Multiple Populations in Globular Clusters: their
role in the Galaxy assembly" (PI: E. Carretta), and from PRIN MIUR 2010-2011,
project \lq{The Chemical and Dynamical Evolution of the Milky Way and Local Group Galaxies}\rq, prot. 2010LY5N2T (PI: F. Matteucci). 
\end{acknowledgements}

\begin{appendix}

\section{The RGB bump of Sculptor}

The RGB bump is produced when the advancing H-burning shell encounters 
the H-abundance discontinuity left over by the outer convection at its maximum depth reached during the first dredge-up. 
The consequent sudden increase of the H-abundance in the shell alters the efficiency of the H-burning shell, and 
causes a temporary drop in the surface luminosity. After the shell has crossed the 
H-abundance discontinuity, the luminosity starts to increase again. 
As a consequence, a low-mass RGB star crosses the same luminosity interval 
three times, and a bump (a local maximum) appears in the star counts per magnitude bin \citep[see, i.e.,][]{iben, salarisbook}.

\begin{figure}
\centering
\includegraphics[width=\columnwidth]{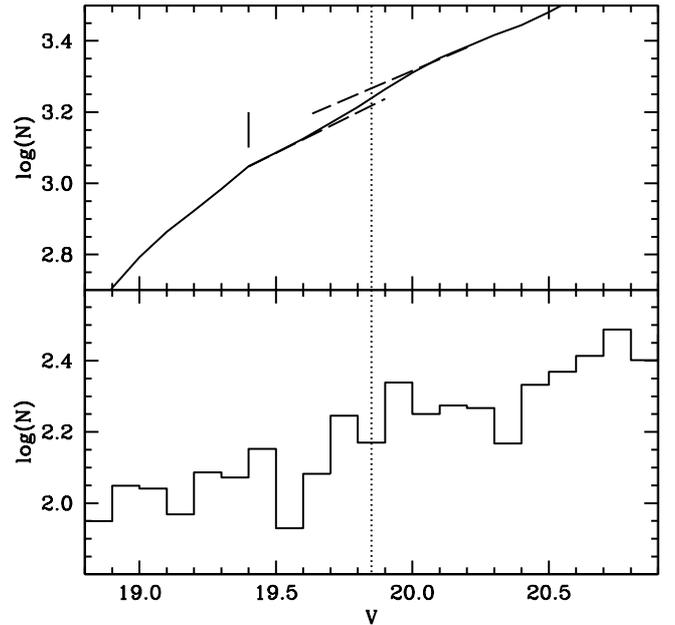}
\caption{Differential (lower panel) and cumulative (upper panel) RGB luminosity functions 
for Sculptor RGB and AGB stars, with ${\rm r_{ell}}<$0.183~deg. The break in the slope of the 
cumulative luminosity function (dashed lines) points to the location of RGB bump, 
whose average magnitude is marked in both panels by vertical dotted lines. The thick solid mark denotes 
a discontinuity in the slope of the cumulative luminosity function, the we ascribe to the appearance of the 
AGB clump (see text for details).}
\label{RGBlf}
\end{figure}

\begin{figure}
\centering
\includegraphics[width=\columnwidth]{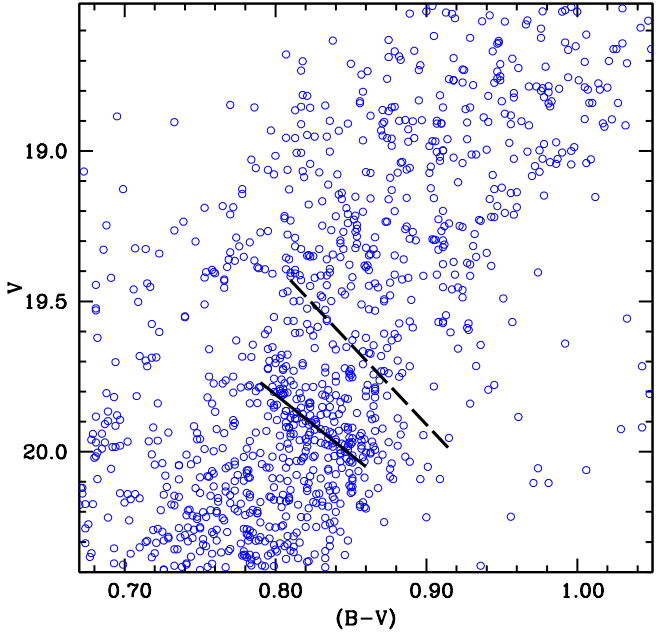}
\caption{Sculptor CMD for RGB stars around the bump (for ${\rm r_{ell}}<$0.183~deg). The solid line 
marks the ridge line of the observed bump region; the dashed line denotes the ridge line of the 
bump for the synthetic sample of RGB stars.}
\label{RGB}
\end{figure}

The analysis of the photometry by \citet{bumpSc} disclosed the presence of two distinct RGB bumps;  
a bright blue bump at V$\sim$19.3-19.4, and a faint red bump one at V$\sim$20.0-20.1. 
Figure~\ref{RGBlf} displays our  
differential and cumulative RGB luminosity functions 
for Sculptor RGB and AGB stars, with ${\rm r_{ell}}<$0.183~deg. The break in the slope of the 
cumulative luminosity function displayed by dashed lines, 
points to the location of the RGB bump \citep{fp:90}. This is strongly corroborated by the CMD of 
Fig.~\ref{RGB}, that displays along the RGB a tilted strip with a strong concentration of stars, 
approximately centred around the magnitude of the 
break in the slope of the luminosity function.
We thus find a single, very extended -- in both V and (B-V) -- and  
continuous bump region whose ridge line is marked in Fig.~\ref{RGB}. The magnitude range 
agrees approximately with the brightness of the fainter bump claimed by \citet{bumpSc}.
The bump gets fainter towards red colours as expected, given that more metal rich populations are redder along the RGB  
and have a fainter bump (at constant age). The spread of the bump over a large range of magnitudes 
makes the feature somewhat not well defined in the differential luminosity function.

We do not find any signature of an additional bump at V$\sim$19.3-19.4, as it is quite clear 
by just examining the CMD, 
although we find a discontinuity in the slope of the cumulative luminosity function 
at V$\sim$19.4 -- approximately the magnitude of the bright bump 
claimed by \citet{bumpSc} --  that we are going to discuss later. 

Figure~\ref{RGB} also displays the ridge line of the bump region from our simulations calculated with the galaxy SFH. 
There is a clear offset of $\sim$0.35~mag in V, where the theoretical is brighter. Notice that there 
is no freedom to shift the simulated CMD to fainter magnitudes, because it would destroy 
the agreement of the synthetic HB with observations.  
Using the simulated CMD as a guide, the colour range of the observed bump corresponds to 
[Fe/H] values ranging between $\sim-$1.5 and $\sim -$2.0.
This disagreement is consistent with the discrepancy found in a sample of Galactic globular clusters,  
in approximately the same metallicity interval, by \citet{dicecco, bumpGC}\footnote{
Synthetic CMDs considering the appropriate SFH for the whole observed area within  ${\rm r_{ell}}=$1~deg, confirm 
that the theoretical RGB bump is systematically brighter than the observed one, and the position of both bumps 
is substantially the same as for the case of ${\rm r_{ell}}<$0.183~deg}. 

%However, \citet{monelli10} compared  
%the difference between the V magnitude of the RGB bump and the HB at the level of the RR Lyrae instability strip 
%(taken at the ZAHB) ${\rm \Delta V_{HB}^{bump}}$ 
%for four dwarf galaxies (Tucana, LGS3, IC1613, Cetus), and     
%determined an average difference (observed-synthetic) of 0.13$\pm$0.14, which is not statistically significant.   
%All these analyses have been performed with the same BaSTI models employed here.

A previous analysis by \citet{monelli10} compared  
the difference between the V magnitude of the RGB bump and the HB at the level of the RR Lyrae instability strip 
(taken at the ZAHB) ${\rm \Delta V_{HB}^{bump}}$ observed in Sculptor, with results from the BaSTI models. 
They used data from the literature that were available at that time, and considered 
two values of ${\rm \Delta V_{HB}^{bump}}$, associated to the two bumps 
found by \citet{bumpSc}. 
As mentioned before, we do not find any trace of the bright bump in our adopted photometry (and there is no 
double RGB bump in the synthetic CMDs either), and we speculate 
that this may correspond to the asymptotic giant branch (AGB) clump, that marks the first 
ignition of the He burning shell around the inert CO core. Essentially AGB stars 
begin to contribute more substantially to the population of red giants from  
the level of the clump, when the evolutionary speed tends to (relatively) 
slow down.  

\begin{figure}
\centering
\includegraphics[width=\columnwidth]{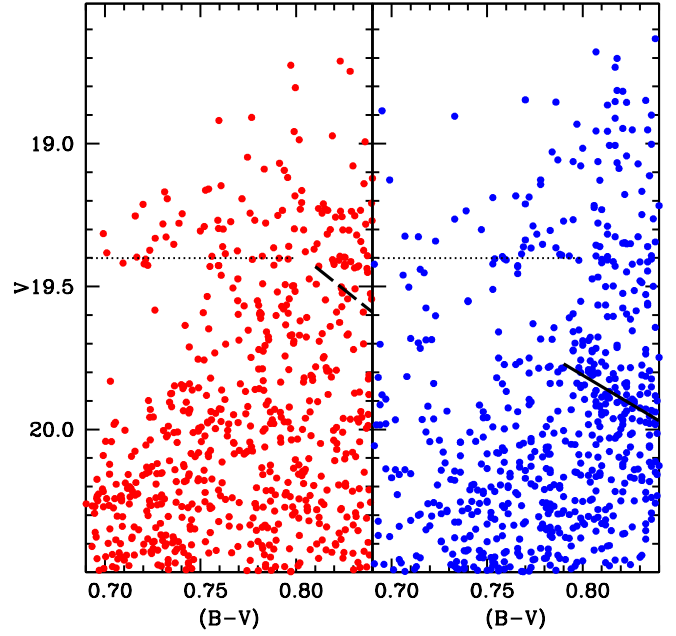}
\caption{Synthetic CMD for stars around the bump and AGB clump (left panel -- SFH for ${\rm r_{ell}}<$0.183~deg), 
compared with observations (right panel). The ridge lines of both observed and synthetic RGB bumps are 
also marked (solid and dashed lines, respectively).  
The AGB clump is the feature on the blue side of the RGB, in the range 
V$\sim$19.1-19.4 and (B-V)$\sim$0.70-0.75. The dotted line marks 
the level of the discontinuity of the slope of the RGB+AGB cumulative luminosity function (see text for details).}
\label{AGB}
\end{figure}

Figure~\ref{AGB} displays a synthetic CMD for the population with ${\rm r_{ell}}$=0.183~deg -- 
and approximately the same number of stars as observed -- compared to the observations. The dotted horizontal line 
marks the magnitude of the discontinuity of the RGB+AGB cumulative luminosity function, brighter than the RGB bump, 
disclosed by Fig.~\ref{RGBlf}.
The AGB clump in the synthetic diagram appears in the region (B-V)$\sim$0.70-0.75, on the blue side of the RGB,  
and V$\sim$19.1-19.4.
\footnote{The clump appears slightly better defined, at the position predicted by the simulations, in the observed CMD of the whole 
region with ${\rm r_{ell}}<$1~deg.} 
The dotted line essentially marks the lower envelope of AGB clump stars and agrees with the magnitude of the brighter and 
bluer bump claimed by \citet{bumpSc}. 

Let's consider the case of the fainter bump, roughly consistent  
with the bump region found in our data, that gives ${\rm \Delta V_{HB}^{bump}}$=$-$0.35$\pm$0.21 
%As mentioned before, the location of the RGB bump in our data is roughly consistent with the faint 
%bump of \citet{bumpSc}. As fo rthe bright one, that we do not find along our RGB data, we 
%suspect it 
%${\rm \Delta V_{HB}^{bump}}$=$-$0.35$\pm$0.21 and ${\rm \Delta V_{HB}^{bump}}$=$-$0.90$\pm$0.21 and. 
This was derived from 
\citet{bumpSc} data, by considering the ZAHB level of the red HB -- that agrees with our CMD --   
and the faint bump, and an average [M/H]$-$1.30$\pm$0.15 from \citet{kir:09}.
With these values and associated error bars, the observed ${\rm \Delta V_{HB}^{bump}}$ appears broadly 
consistent with theoretical models, for ages between 10 and 14~Gyr \citep[see Fig.5 of][]{monelli10}.
Our analysis suggests instead that the theoretical ${\rm \Delta V_{HB}^{bump}}$ is smaller than observations (the theoretical bump 
is too bright with respect to the HB level, taken as reference).

The difference with our result may be traced back to a combination of factors. 
First of all, [Fe/H] for the stars that define the bump region, turns out to lie in the range 
${\rm -2.0 < [Fe/H] < -1.5}$. 
The [$\alpha$/Fe] values from the SFH, combined with these [Fe/H], give a range of 
[M/H] values lower than the [M/H]$-$1.30$\pm$0.15 assumed by \citet{monelli10}, and hence a brighter theoretical bump.
In second instance, our HB simulations show that for the red HB (stars within the {\sl red} box in our analysis) 
the ZAHB level, taken as lower envelope of the observed stellar distribution, is determined by the more metal rich population, 
at variance with the metallicity range that dominates the bump region.
Both these factors go in the direction of reducing the theoretical ${\rm \Delta V_{HB}^{bump}}$ to be used against the observed 
value, compared to the theoretical values in \citet{monelli10} study. 

\end{appendix}

%\bibliographystyle{aa}
%\bibliography{Sculptor_HB}

\end{document}